\documentclass[aps,prl,showpacs,nofootinbib,twocolumn,superscriptaddress]{revtex4}
\usepackage{epsf}
\usepackage{graphicx}
\usepackage{amsmath}
\def\slashchar#1{\setbox0=\hbox{$#1$}
   \dimen0=\wd0 \setbox1=\hbox{/} \dimen1=\wd1
   \ifdim\dimen0>\dimen1 \rlap{\hbox to \dimen0{\hfil/\hfil}} #1
   \else  \rlap{\hbox to \dimen1{\hfil$#1$\hfil}} / \fi}

\begin{document}

\title{$N$-$\Delta(1232)$ axial form factors from
  weak pion production} 

\author{E. Hern\'andez} \affiliation{Departamento de F\'\i sica Fundamental 
e IUFFyM,\\ Universidad de Salamanca, E-37008 Salamanca, Spain} 
\author{J.~Nieves}
\affiliation{Instituto de F\'\i sica Corpuscular (IFIC), Centro Mixto
CSIC-Universidad de Valencia, Institutos de Investigaci\'on de
Paterna, Aptd. 22085, E-46071 Valencia, Spain} 
\author{M.~Valverde}
\affiliation{Research Center for Nuclear Physics (RCNP), Osaka
University, Ibaraki 567-0047, Japan} 
\author{M.J. \surname{Vicente Vacas}}
\affiliation{Departamento de F\'\i sica Te\'orica and IFIC, Centro Mixto
Universidad de Valencia-CSIC, Institutos de Investigaci\'on de
Paterna, Aptd. 22085, E-46071 Valencia, Spain}

\pacs{25.30.Pt,13.15.+g}

\begin{abstract}
The $N\Delta$ axial form factors are determined from neutrino induced
pion production ANL \& BNL data by using a state of the art
theoretical model, which accounts both for background mechanisms and
deuteron effects. We find violations of the off diagonal
Goldberger-Treiman relation at the level of $2\sigma$ which might have
an impact in background calculations for T2K and MiniBooNE low energy
neutrino oscillation precision experiments.
\end{abstract}

\maketitle


The $\Delta$(1232) resonance is the lightest baryonic excitation of
the nucleon. In addition, it couples very strongly to the lightest
meson, the pion, and to the photon. As a consequence, the
$\Delta$(1232) is of the utmost importance in the description of a
wide range of hadronic and nuclear phenomenology going from low and
intermediate energy processes~\cite{Brown:1975di,Cattapan:2002rx} to
the GZK cut-off of the cosmic ray
flux~\cite{Greisen:1966jv,Zatsepin:1966jv}. On the other hand, despite
its large width, it is well separated from other resonances what
facilitates its experimental investigation. In particular, the
electromagnetic nucleon to $\Delta$(1232) excitation processes,
induced by electrons and photons, have been extensively studied at
many experimental facilities like LEGS, BATES, ELSA, MAMI, and
J-LAB. For a recent review see Ref.~\cite{Pascalutsa:2006up}, where
also many of the recent theoretical advances in the understanding of
the resonance have been addressed.

There has also been a great theoretical interest in the axial nucleon
$\Delta$ transition form factors. Recently, they have been studied
using quark models \cite{BarquillaCano:2007yk}, Light Cone QCD Sum
Rules~\cite{Aliev:2007pi}, Lattice QCD~\cite{Alexandrou:2006mc} and
Chiral Perturbation Theory
($\chi$PT)~\cite{Geng:2008bm,Procura:2008ze}.  These form factors
are of topical importance in the background analysis of some of the
neutrino oscillation experiments (e.g.~\cite{AguilarArevalo:2008rc}).
However, their experimental knowledge  is less than
satisfactory. Although the feasibility of their extraction in
parity-violating electron scattering has been
considered~\cite{Mukhopadhyay:1998mn}, the best available information
comes from old bubble chamber neutrino scattering experiments at
ANL~\cite{Barish:1978pj,Radecky:1981fn} and
BNL~\cite{Kitagaki:1986ct,Kitagaki:1990vs}. These experiments measured
pion production in deuterium at relatively low energies where the
dominant contribution is given by the $\Delta$ pole ($\Delta P$)
mechanism: weak excitation of the $\Delta(1232)$ resonance and its
subsequent decay into $N\pi$.  Only very recently, $\pi^0$ production
cross sections have been measured at low neutrino energies and with
good statistics~\cite{AguilarArevalo:2009ww}.  However, the target was
mineral oil what implies large and difficult to disentangle nuclear
effects. Thus, these data are less well suited for the extraction of
the $N\Delta$ axial form factors.

Besides the original experimental publications, there are many studies
of the ANL and/or the BNL data in the
literature~\cite{AlvarezRuso:1998hi,Hernandez:2007qq,Pa04,Pa05,Leitner:2008ue,Graczyk:2009qm}
with different advantages and shortcomings. Some of those studies are discussed
below. In this letter, we analyze the ANL and BNL data incorporating
the deuteron effects, with a proper consideration of statistical and
systematical uncertainties and taking advantage of several recent
developments: improved vector form factors and a new model for weak
pion production off the nucleon that includes background terms.

A convenient parameterization of the $W^+n\to\Delta^+$ vertex is given
in terms of eight $q^2$ (momentum transfer square) dependent form-factors: four vector and four
axial ($C^A_{3,4,5,6}$ ) ones. We follow the conventions and notation
of Ref.~\cite{Hernandez:2007qq}. Vector form factors have been
determined from the analysis of photo and electro-production
data. Here, we use the parameterization of Lalakulich {\it et
  al.}~\cite{Lalakulich:2006sw}, as done in
Ref.~\cite{Hernandez:2007qq}.  Among the axial form factors the most
important contribution comes from $C_5^A$. The form factor $C_6^A$,
which contribution to the differential cross section vanishes for
massless leptons, can be related to $C_5^A$ thanks to the partial
conservation of the axial current ($ C_6^A(q^2) = C_5^A(q^2)
\frac{M^2}{m_\pi^2-q^2}$, with $m_\pi$ and $M$ the pion and nucleon
masses, respectively). Since there are no other theoretical
constraints for $C_{3,4,5}^A(q^2)$, they have to be fitted to data.
Most analysis, including the ANL and BNL ones, adopt Adler's
model~\cite{adler} where\footnote{Setting $C_3^A$ to zero seems to be
  consistent with SU(6) symmetry~\cite{Liu:1995bu} and recent lattice
  QCD results~\cite{negele07-prd}.}  $C_3^A(q^2) = 0$ and $C_4^A(q^2)
= -{C_5^A(q^2)}/{4}$.  For $C_{5}^A$ several $q^2$ parameterizations
have been used~\cite{Pa04, Leitner:2008ue}, though given the limited
range of statistically significant $q^2$ values accessible in the ANL
and BNL data, it should be sufficient to consider for it a dipole
dependence, $C_5^A(q^2) = \frac{C_5^A(0)}{(1-q^2/M^2_{A\Delta})^2}$,
where one would expect $M_{A\Delta}\sim 0.85-1$ GeV, to guarantee an
axial transition radius\footnote{ It is defined from
  $C_5^A(q^2)/C_5^A(0) = 1+q^2 R_A^2/6+{\cal O}(q^4).$} $R_A$ in the
range of $0.7-0.8$ fm, and $C_5^A(0)\sim 1.2$, which is the prediction
of the off-diagonal Goldberger-Treiman relation (GTR),
$C_5^A(0)=\sqrt{\frac23}f_\pi\frac{f^*}{m_\pi}=1.2$, with the $\pi
N\Delta$ coupling $f^* = 2.2$ fixed to the $\Delta$ width and
$f_\pi\sim 93$ MeV, the pion decay constant.

There is no constraint from $\chi$PT and lattice calculations are still
not conclusive about the size of possible violations of the GTR. For instance,
though values for $C_5^A(0)$ as low as 0.9 can be inferred in the
chiral limit from the results of Ref.~\cite{negele07-prd}, they also predict 
$C_5^A(0)/\left(\sqrt{\frac23}f_\pi\frac{f^*}{m_\pi}\right)$ to be
greater than one.

\section{$C_5^A(q^2)$ assuming $\Delta P$ dominance} 

Traditionally, Adler's model and the GTR have been assumed, being the
$M_{A\Delta}$ axial mass adjusted in such a way that the $\Delta P$
contribution alone would lead to a reasonable description of the shape
of the BNL $q^2$ differential $\nu_\mu p \to \mu^- p \pi^+$ cross
section (see e.g. Ref.~\cite{Pa04}). These fits also describe
reasonably well the $q^2$ dependence of the ANL data and the BNL total
cross section but overestimate the size of the ANL data by 20\% near
the maximum~\cite{Pa05}. Thus, ANL data might favor $C_5^A(0)$ values
smaller than the GTR prediction.

Recently, two re-analysis have been carried out trying to make
compatible the GTR prediction for $C_5^A(0)$ and ANL data. In
Ref.~\cite{Leitner:2008ue}, $C_5^A(0)$ is kept to its GTR value and
three additional parameters, that control the $C_5^A(q^2)$ fall off,
are fitted to the ANL data. In fact $C_5^A(q^2\sim 0)$ is not so
relevant due to phase space, and what is actually important is the
$C_5^A(q^2)$ value in the region around $-q^2\sim
0.1\,$GeV$^2$. Although ANL data are well reproduced, we find the
outcome in ~\cite{Leitner:2008ue} to be unphysical, because it
provides a quite pronounced $q^2-$dependence that gives rise to a too
large axial transition radius\footnote{Further details and possible
repercussions in neutrino induced coherent pion production
calculations are discussed in \cite{Hernandez:2009zg}. There, ANL
data fits of the type proposed in \cite{Leitner:2008ue}, but including
chiral non-resonant contributions are also performed, finding that
then the axial transition radius becomes even larger, about 2.5 fm.}
of around 1.4 fm. Moreover, neither the fitted parameter statistical
errors, nor the corresponding correlation coefficients are calculated
in ~\cite{Leitner:2008ue}. Undoubtedly, the fit carried out there
should be quite unstable, from the statistical point of view, because
of the difficulty of determining three parameters given the limited
range of $q^2$ values covered in the ANL data set. Furthermore, the
consistency of these results with the BNL data has not been tested.

A second re-analysis~\cite{Graczyk:2009qm} brings in the discussion
two interesting points. First that both ANL and BNL data were measured
in deuterium, and second, the uncertainties in the neutrino flux
normalization.  Deuteron structure effects in the $\nu d \to
\mu^-\Delta^{++} n$ reaction, sometimes ignored, were estimated from
the results of Ref.~\cite{AlvarezRuso:1998hi} to produce a reduction 
of the cross section from 5--10\%.  In what respects to the ANL and BNL flux uncertainties,
the procedure followed in \cite{Graczyk:2009qm} is not robust from the
statistical point of view, since it ignores the correlations of these
systematic errors\footnote{There exist some other aspects that might
require further investigation. For instance, additional parameters
$p_{\rm ANL}$ and $p_{\rm BNL}$ are introduced in
\cite{Graczyk:2009qm} (see $\chi^2$ function in Eq.~(37)) to account
for the flux uncertainties. At very low $q^2$ values, $d\sigma/dq^2$
is totally dominated by $C_5^A$. If we had infinitely precise
statistical measurements, the fit carried out in \cite{Graczyk:2009qm}
would provide a very precise determination of the ratio
$C_5^A(0)/\sqrt p$, but not of the form factor $C_5^A(0)$. However, in
such situation, one expects to extract $C_5^A(0)$, though with an
uncertainty dominated by that of the neutrino flux
normalization. Besides, the fit to the BNL data uses the total
cross-section data, for which the hadronic invariant mass is
unconstrained, and the neutrino energy varies in the range 0.5--3
GeV. Above 1 GeV, heavier resonances than the $\Delta(1232)$, and not
considered in~\cite{Graczyk:2009qm}, should play a
role~\cite{Lalakulich:2006sw}.}. Nevertheless, this latter work
constitutes a clear step forward, and from a combined best fit to the
ANL \& BNL data, the authors of \cite{Graczyk:2009qm} find
$C_5^A(0)=1.19 \pm 0.08$ in agreement with the GTR estimate.

\section{Axial form factors including 
the chiral non-resonant background.}

All the above-mentioned determinations of $C_5^A(q^2)$ suffer from a
serious theoretical limitation. Though the $\Delta P$ mechanism
dominates the neutrino pion production reaction, specially in the
$\Delta^{++}$ channel, there exist sizable non-resonant contributions
of special relevance for low neutrino energies (below 1 GeV) of
interest in T2K and MiniBooNE experiments.  These background terms are
totally fixed by the pattern of spontaneous chiral symmetry breaking
of QCD, and are given in terms of the nucleon and pion masses, the
axial charge of the nucleon and the pion decay constant. When
background terms are considered, the tension between ANL data and the
GTR prediction for $C_5^A(0)$ substantially increases. Indeed, the fit
carried out in ~\cite{Hernandez:2007qq} to the ANL data finds a value
for $C_5^A(0)$ as low as $0.87 \pm 0.08$ with a reasonable axial
transition radius of $0.75 \pm 0.06$ fm, and a large Gaussian correlation
coefficient ($r=0.85$), as expected from the above discussion of the
results of Ref.~\cite{Leitner:2008ue}.

Here, we follow the approach of Ref.~\cite{Hernandez:2007qq}, but
implementing four major improvements: {\it i)} we include in the fit
the BNL total $\nu_\mu p \to \mu^- p \pi^+$ cross section measurements
of Ref.~\cite{Kitagaki:1986ct}. Since there is no cut in the outgoing
pion-nucleon invariant mass in the BNL data, and in order to avoid
heavier resonances from playing a significant role, we have just
included the three lowest neutrino energies: 0.65, 0.9 and 1.1 GeV.
We do not use the BNL measurement of the $q^2-$differential cross
section, since it lacks an absolute normalization.  {\it ii)} we take
into account deuteron effects in our theoretical calculation, {\it
iii)} we treat the uncertainties in the ANL and BNL neutrino flux
normalizations as fully correlated systematic errors, improving thus
the treatment adopted in Ref.~\cite{Graczyk:2009qm}, and finally {\it
iv)} in some fits, we relax the Adler's model constraints, by setting
$C_{3,4}^A(q^2)= C_{3,4}^A(0)\,( C_{5}^A(q^2)/C_{5}^A(0))$, and
explore the possibility of extracting some direct information on
$C_{3,4}^A(0)$.

Let us consider first the neutrino--deuteron reaction
$\nu d \to \mu^-p\pi^+ n$ measured in ANL and
BNL.  Owing to the inclusion of background terms, the formalism of
Ref.~\cite{AlvarezRuso:1998hi}, where the $p\pi^+$ pair was replaced
by a $\Delta^{++}$, cannot be used to account for deuteron
corrections, and we must work with four particles in the final
state. Neglecting the $D-$wave deuteron component and considering the
neutron as a mere spectator, we find for the differential cross section
on deuteron
\begin{eqnarray}
\frac{d\sigma}{dq^2 dW}\Big|_d = \int d^3p_d |\Psi_d(\vec{p}_d)|^2
\frac{M}{E_{p,d}} \frac{d\sigma}{dq^2 dW}\Big|_{p-{\rm offshell}}
\end{eqnarray}
where $E_{p,d}=m_d-\sqrt{M^2+\vec{p}_d^{~2}}$, with $m_d$ the deuteron
mass, is the energy of the off-shell proton inside the deuteron which
has four-momentum $p^\mu = \left (E_{p,d}, \vec{p}_d\right)$.  $W$ is
the final $p\pi^+$ invariant mass.  The differential cross section
$d\sigma/dq^2 dW\Big|_{p-{\rm offshell}}$ is computed using the model
of Ref.~\cite{Hernandez:2007qq}. Finally, $\Psi_d$ is the $S-$wave
Paris potential deuteron wave function~\cite{Lacombe:1981eg}
normalized to 1.

In what respects to the neutrino flux normalization uncertainties, we
consider them as sources of 20\% and 10\% systematic errors for the
ANL and BNL experiments respectively (see discussion in
~\cite{Graczyk:2009qm}). We have assumed that the ANL and BNL input
data have independent statistical errors ($\sigma_i$) and
fully-correlated systematic errors ($\epsilon_i$), but no correlations
linking the ANL and BNL sets. We end up with a $12\times 12$
covariance matrix, $C$, with two diagonal blocks. The first $9\times
9$ block is for the ANL flux averaged $q^2-$differential $\nu d \to
\mu^-p\pi^+n$ cross section data (with a 1.4 GeV cut in $W$), while
the second $3\times 3$ block is for the BNL total cross sections
mentioned above. Both blocks have the form $C_{ij}=\sigma_i^2
\delta_{ij}+\epsilon_i\epsilon_j$. The $\chi^2$ function is
constructed by using the inverse of the covariance matrix. 

\begin{figure*}[tbh]
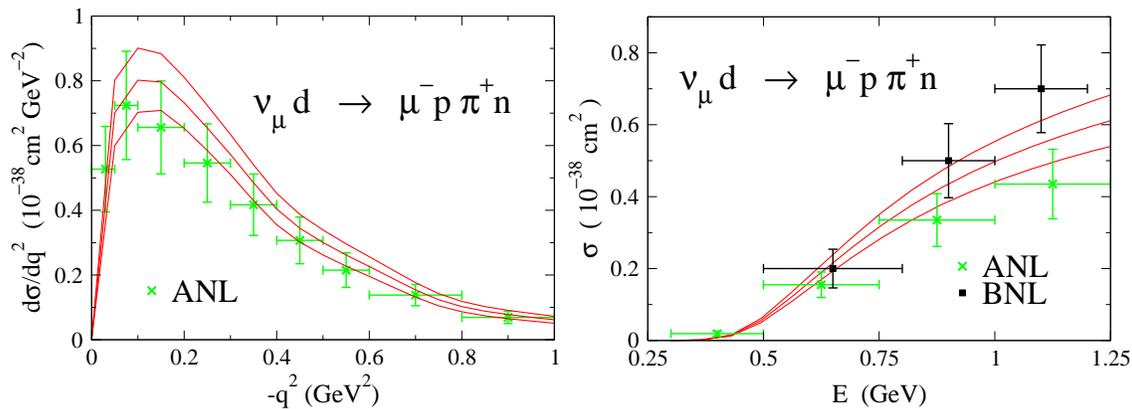

\begin{center}
\makebox[0pt]{\includegraphics[scale=0.285]{dq2_ANL.eps}\hspace{.01cm}
\includegraphics[scale=0.285]{ener.eps}}\\
\end{center}
\caption{\footnotesize Comparison of the ANL $d\sigma/dq^2$ differential
(left panel) and  ANL~\&~BNL total (right panel) cross section data with
fit IV theoretical results. Theoretical 68\% confidence
level bands are also displayed. Data in both plots include a systematic error (20\%
for ANL and 10\% for BNL data) added in quadrature to the statistical
ones. In the left panel, both data and results  include a cut $W<1.4\,$GeV.
}
\label{fig:res}
\end{figure*}

Results
from several fits are compiled in Table~\ref{tab:fit}, from where we 
draw several conclusions. First, by comparing fit II$^*$ with
Ref.~\cite{Hernandez:2007qq}, we deduce that the consideration of BNL
data and flux uncertainties increases the value of $C_5^A(0)$ by about
9\%, while strongly reduces the statistical correlations between
$C_5^A(0)$ and $M_{A\Delta}$. Second, the inclusion of background
terms reduces $C_5^A(0)$ by about 13\%, while deuteron effects increase
it by about 5\%, consistently with the results of
\cite{Hernandez:2007qq} and \cite{AlvarezRuso:1998hi,Graczyk:2009qm},
respectively. Third, the fitted data are quite insensitive to
$C_{3,4}^A(0)$, as fit V--VII results show. This is easily understood,
taking for simplicity the massless lepton limit. In that case
\begin{equation}
\frac{d\sigma}{dq^2} \propto \left \{ [C_5^A(0)]^2 + q^2 a(q^2) \right \}
\end{equation}
and $C_{3,4}^A(0)$ start contributing to $a(q^2)$, i.e. to ${\cal
O}(q^2)$, which also gets contributions from vector form factors and
terms proportional to $dC_5^A/dq^2\big|_{q^2=0}$. This also explains
the large statistical correlations displayed in fits V--VII. Moreover,
$dC_{3,4}^A/dq^2\big|_{q^2=0}$ appears at order ${\cal O}(q^4)$, which
has prevented us to fitting the $q^2-$shape of these form
factors. Fourth, fit IV is probably the most robust from the
statistical point of view. In Fig.~\ref{fig:res}, we display fit IV
results for the ANL and BNL $\nu d \to \mu^-p\pi^+ n$ cross sections.
Looking at the central values of $C_5^A(0)$, we conclude that the
violation of the off-diagonal GTR is about 15\% smaller than that
suggested in Ref.~\cite{Hernandez:2007qq}, though it is definitely
greater than that claimed in \cite{Graczyk:2009qm}, mostly because in
this latter work background terms were not considered. However, GTR
and fit IV $C_5^A(0)$ values differ in less than two sigmas, and the
discrepancy is even smaller if Adler's constraints are removed.  These
new results are quite relevant for the neutrino induced coherent pion
production process in nuclei which is much more forward peaked than
the incoherent reaction.  For instance, we expect the results in
Ref.~\cite{Amaro:2008hd}, based in the determination of $C_5^A(0)$ of
Ref.~\cite{Hernandez:2007qq}, to underestimate cross sections by at
least 30\%.

By using a state of the art theoretical model, we have determined the
$N\Delta$ axial form factors from statistically improved fits to the
combined ANL \& BNL data. The inclusion of chiral background terms
significantly modifies the form factors. We have found violations of
the GTR at the level of $2\sigma$, when the usual Adler's constraints
are adopted. This will influence background calculations for T2K and
MiniBooNE low energy neutrino precision oscillation experiments.

\begin{center}
\begin{table*}
\caption{Results from different fits to the ANL and BNL
  data. Deuteron effects are included in all cases except for the two
  first fits (marked with $*$). The non-resonant chiral background
  contributions are not included in fits I and III. In the
  $C_{3,4}^A$ columns, {\it Ad} indicates that  Adler's
  constraints ($C_3^A=0,\;C_4^A=-C_5^A/4$) are imposed. Finally, $r_{ij}$
  are Gaussian correlation coefficients between parameters $i$ and
  $j$. For $C_5^A(q^2)$ a dipole form has been used.}
\begin{tabular}{c|cccc|cccccc|c}
\hline\hline
& $C_5^A(0)$ & $M_{A\Delta}$/GeV & $C_3^A(0)$ & $C_4^A(0)$& $r_{12}$
  & $r_{13}$ & $r_{14}$ & $r_{23}$ & $r_{24}$ & $r_{34}$ &$ \chi^2$/dof\\\hline  
I$^*$ (only $\Delta P$) &  $1.08\pm 0.10$ & $0.92\pm 0.06$& Ad & Ad &
  $-0.06$ &&&&&& 0.36\\
II$^*$ &  $0.95\pm 0.11$ & $0.92\pm 0.08$& Ad & Ad &
  $-0.08$ &&&&&& 0.49\\ \hline
III (only $\Delta P$) &  $1.13\pm 0.10$ & $0.93\pm 0.06$& Ad & Ad &
  $-0.06$ &&&&&& 0.32\\
IV &  $1.00\pm 0.11$ & $0.93\pm 0.07$& Ad & Ad &
  $-0.08$ &&&&&& 0.42 \\
 V &  $1.08\pm 0.14$ & $0.91\pm 0.10$& $-1.0\pm 1.4$ & Ad &
  $-0.48$ &$-0.61$&&$0.81$&&& 0.40 \\
 VI &  $1.08\pm 0.14$ & $0.86\pm 0.15$& Ad & $-1.0\pm 1.3$ &
  $-0.57$ &$-0.66$&&$0.93$&&& 0.40 \\
VII &  $1.07\pm 0.15$ & $1.0\pm 0.3$& $1\pm 4$ & $-2\pm 4$ &
  $-0.62$ &$-0.45$&0.30&$0.89$&$-0.77$&$-0.97$& 0.44 \\
\hline\hline
\end{tabular}
\label{tab:fit}
\end{table*}
\end{center}

\begin{acknowledgments}
M.Valverde acknowledges the Japanese Society for the Promotion of Science
(JSPS) for a Postdoctoral Fellowship.  Research supported by DGI
contracts FIS2008-01143, FIS2006-03438, FPA2007-65748 and
CSD2007-00042, JCyL contracts SA016A07 and GR12, Generalitat
Valenciana contract PROMETEO/2009/0090 and by EU HadronPhysics2
contract 227431.
\end{acknowledgments}

\end{document}